\documentclass[%
 reprint,
 amsmath,amssymb,
 aps,
]{revtex4-1}

\newcommand{\Pib}{\mbox{\boldmath $\Pi $}}

\newcommand{\rhob}{\mbox{\boldmath $\rho $}}

\newcommand{\bk}{{\bf k}}
\newcommand{\be}{{\bf e}}

\newcommand{\bB}{{\bf B}}
\newcommand{\bq}{{\bf q}}

\newcommand{\bp}{{\bf p}}

\newcommand{\br}{{\bf r}}

\newcommand{\bR}{{\bf R}}
\newcommand{\bP}{{\bf P}}

\newcommand{\beq}{\begin{equation}}
\newcommand{\beqn}{\begin{eqnarray}}
\newcommand{\eeq}{\end{equation}}
\newcommand{\eeqn}{\end{eqnarray}}

\usepackage{graphicx}
\usepackage{dcolumn}
\usepackage{bm,color}

\begin{document}


\title{Charged exctions in two-dimensional transition-metal dichalcogenides -- semiclassical calculation of Berry-curvature effects}

\author{A. Hichri}
 
 \altaffiliation[Also at ]{Laboratoire de Physique des Matériaux, Facult\'e des Sciences de Bizerte, Universit\'e de Carthage, 7021 Zarzouna, Tunisie.}
\author{ S. Jaziri}%
 \email{Sihem.Jaziri@fsb.rnu.tn}
\affiliation{ Laboratoire de Physique des Matériaux, Facult\'e des Sciences de Bizerte, Universit\'e de Carthage, 7021 Zarzouna, Tunisie
}%

\affiliation{Laboratoire de Physique de la Matière Condens\'ee, Facult\'e des Sciences de Tunis, Universit\'e de Tunis El Manar, 2092 El Manar, Tunisie}%
\author{ M. O. Goerbig}%
\affiliation{
Laboratoire de Physique des Solides,  CNRS UMR
8502, Universit\'e Paris-Sud, Univerist\'e Paris Saclay, 91405 Orsay Cedex, France
}%





\date{\today}

\begin{abstract}

We theoretically study the role of the Berry curvature on neutral and charged excitons in two-dimensional transition-metal dichalcogenides. The Berry curvature
arises due to a strong coupling between the conduction and valence bands in these materials that can to great extent be described within the model
of massive Dirac fermions. The Berry curvature lifts the degeneracy of exciton states with opposite angular momentum. 
Using an electronic interaction that accounts for non-local screening effects, we find a Berry-curvature induced splitting of $\sim 17$ meV between the 2\textit{p}$_{-}$  
and 2\textit{p}$_{+}$ exciton states in WS$_2$, consistent with experimental findings. Furthermore, we calculate the trion binding energies in WS$_2$ and WSe$_2$ for 
a large variety of screening lenghts and different dielectric constants for the environment. Our approach indicates the prominent role played by the Berry curvature 
along with non-local electronic interactions in the understanding of the energy spectra of neutral and charged excitons in transition-metal dichalcogenides and in the 
the interpretation of their optical properties. 

\end{abstract}

\maketitle


\section{Introduction}

Monolayer transition-metal dichalcogenides (TMDC) are a particular class of two-dimensional (2D) materials that exhibit massive Dirac fermions at the two inequivalent valleys 
$K$ and $K'$ in the first Brillouin zone \cite{Xiao2012, Xiao2013,Xiao2014,Xu2014,Yu2014,Glasov2014, MacDonald2015,Xu2016,Stier2016,Borghardt2017,Wang2018,Molas2018}. Beyond graphene, they are therefore a fascinating condensed-matter platform for studying relativistic quantum mechanics in 
low spatial dimensions. A prominent physical phenomenon, where these relativistic effects are expected to occur, is the formation of excitons. Indeed, in several undoped 2D TMDC,
such as MoS$_2$, MoSe$_2$, WS$_2$, or WSe$_2$, the Fermi level is situated in the direct band gap, and an electron promoted to the conduction band is strongly bound to the hole 
left behind in the valence band to form a neutral exciton. Very soon, relatively strong exciton binding energies were found experimentally \cite{Chernikov2014,He2014,Hanbicki2015,Gupta2017}. 

Contrary to excitons in many 2D and 3D materials, the experimentally obtained spectrum could only be poorly fitted by the 2D hydrogen model \cite{Chernikov2014,Ye2014}. A possible source for 
this discrepancy was identified in the form of a particular non-local interaction potential due to complex screening effects in the layered material \cite{Palacios2014,Stier2016,Crooker2016,Wang2018}. While such
screening effects are likely to play a role in the quantitative understanding of the exciton spectra, a qualitatively new perspective was proposed by Zhou \textit{et al.}
\cite{Zhou2015} and Srivastava and Imamoglu \cite{srivastava2015} who pointet out prominent band coupling effects. Even if the electron resides in the conduction and the hole
in the valence band, the bands are two faces of the same medal, e.g. within a description of a massive Dirac model -- the projection to a single band is then accompanied by
additional Berry-curvature terms. These terms consist of a momentum that is coupled to local electric fields generated by the Coulomb interaction between the electron and the hole. 
They yield corrections to the exciton spectrum that are on the order of a $\alpha^2$, where the coupling constant $\alpha=\sqrt{\Omega}/a_B$ is the ratio between the Berry curvature 
$\Omega=\hbar^2/\Delta \mu$ at the direct gap $\Delta$ ($\mu$ is the reduced exciton mass) and the effective Bohr radius $a_B=\hbar^2\kappa/\mu e^2$, in terms of the dielectric
constant $\kappa$ of the environment \cite{Trushin2018}. While these corrections are negligible in large-gap systems, since $\alpha\propto 1/\sqrt{\Delta}$, they are expected
to be pertinent in 2D TMDC, where $\alpha\sim 1$, namely in the $ns$ excitonic states due to there strong decrease with distance \cite{Trushin2018}.

In the present paper, we discuss these band-coupling effects on charged excitons (trions). These trions are formed when the semiconducting material is slightly doped, in which case 
electron-electron interactions favor bound states between the additional charges and neutral excitons. 
Similarly to neutral excitons that are also briefly revisited here, we show that the
Berry-curvature corrections yield a lower binding energy as compared to the case where they are neglected. As for neutral excitons, this is due to the short-range repulsive 
character of the terms including the Berry curvature. Our calculations are performed for interaction potentials that include non-local screening effects in order to allow for a
quantitative comparison with numerical and experimental data. 

The paper is organized as follows. In Sec. \ref{sec:2}, we revisit neutral excitons, starting from a quantum-mechanical approach to include Berry-curvature corrections at linear
order in the one-particle Hamiltonian (Sec. \ref{sec:2.1}) and the exciton Hamiltonian (Sec. \ref{sec:2.2}). Section \ref{sec:trions} is then devoted to a generalization of 
the approach to charged excitons (trions). The formal aspects are presented in Sec. \ref{sec:3.1} for the semiclassical Berry-curvature term and in Sec. \ref{sec:3.2} for the
Darwin term, while Sec. \ref{sec:3.3} is devoted to a quantitative study of the trion binding energies. 

\section{Quantum Hamiltonian of neutral excitons in the presence of band coupling}\label{sec:2}

In the present section, we review the theoretical approach proposed by Zhou \textit{et al.} \cite{Zhou2015} and Srivastava and Imamoglu \cite{srivastava2015} 
to take into account corrective terms in the excitonic Hamiltonian that arise due to a non-zero Berry-curvature. It serves us also for the generalization of these terms
to trions in Sec. \ref{sec:trions}.

\subsection{Single-particle considerations}\label{sec:2.1}

Let us consider the Hamiltonian of a Bloch electron restricted to a single band $\alpha$
\begin{equation}\label{eq:H0}
 H=H_\alpha(\bk) + V(\br),
\end{equation}
where $H_\alpha(\bk)$ is simply the energy of the band we are interested in (be it an electron or a hole band), as a function of the wave vector $\bk$,
and $V(\br)$ is an external potential that varies smoothly on the 
lattice scale. Notice that this description is problematic in the sense that 
the position operator $\br$ necessarily mixes states of other bands that we want to discard in the low-energy 
model. In this sense $\bk$ and $\br$ are not canonical quantum variables but rather (gauge-invariant) semi-classical variables or averaged quantities projected to a single band
$\alpha$. This leads to the semiclassical equations of motion for Bloch electrons and the introduction of the Berry curvature, which modifies the electron's velocity \cite{NiuRev}.

Instead of using the above-mentioned semiclassical equations of motion, we try to construct here a quantum Hamiltonian that yields, to lowest order in the magnetic field and
the Berry curvature, the same equations of motion. Since the Berry curvature plays the role of a magnetic field in reciprocal space, we introduce the gauge-invariant momentum
and position operators 
\begin{equation}\label{eq:peierls}
 \bp \rightarrow \Pib=\bp + \frac{e}{2}\bB\times\br \qquad \br \rightarrow \bR=\br + \frac{1}{2\hbar} \Omega_\alpha\times \bp,
\end{equation}
in a Peierls-type approach, where $B$ is the magnetic field and $\Omega_\alpha$ the Berry curvature of the band $\alpha$ directed in the direction perpendicular to the 2D plane
and $-e$ is the charge of an electron. 
With the help of the usual commutation relations $[x_j,p_{j'}]=i\hbar\delta_{j,j'}$ for the canonical variables $\br$ and $\bp$, one obtains
\begin{equation}\label{eq:comm}
 [\Pi_x,\Pi_y]= -ie\hbar B\qquad [X,Y]= i\Omega_{\alpha},
\end{equation}
in agreement with the semiclassical equations of motion. The quantum Hamiltonian is thus obtained from Eq. (\ref{eq:H0}) with the help of the substitution (\ref{eq:peierls}),
\begin{equation}\label{eq:ham}
 \hat{H} = \hat{H}_0\left(\frac{\bp + e \bB\times \br/2}{\hbar}\right) + \hat{V}\left(\br + \frac{1}{2\hbar}\Omega_{\alpha}\times \bp \right).
\end{equation}
To justify the consistency of this quantum Hamiltonian, we retrieve the semiclassical equations of motion from the Heisenberg equations of motion
\begin{equation}
 i\hbar \dot{X}_\mu=[X_\mu,\hat{H}] \qquad i\hbar \dot{\Pi}_{\mu} =[\Pi_\mu,\hat{H}]
\end{equation}
and the commutator
\begin{eqnarray}\label{eq:comm2}
\nonumber
[X_{\mu},\Pi_{\mu'}] &=& [x_{\mu} + \epsilon_{\mu\nu\sigma}(\Omega_\alpha)_{\nu} p_{\sigma}/2\hbar, p_{\mu'} + e\epsilon_{\mu'\nu'\sigma'}B_{\nu'}x_{\sigma'}/2]\\
&\simeq & i\delta_{\mu,\mu'}(\hbar + e\Omega_\alpha B) = i\tilde{h}\delta_{\mu,\nu},
\end{eqnarray}
to lowest order in the $B$ and $\Omega_\alpha$. In the last line we have introduced a ``deformed'' Planck constant $\tilde{h}=\hbar + eB\Omega_\alpha$ --
this is due to a modified density of states, but arises only at second order (product of $B\Omega_\alpha$). 
Since we are interested, here, only in corrections at order one (and in addition in the 
$B=0$ case), we will set $\tilde{h}=\hbar$ at the end. With this help, one finds from the Heisenberg equations of motion
\begin{eqnarray}
\nonumber\label{eq:heis}
 \dot{\bR} &= & \left(1+\frac{eB\Omega_{\alpha}}{\hbar}\right)\frac{\partial \hat{H}_0}{\partial \Pib} + \frac{\partial \hat{V}}{\hbar\partial \bR}\times\Omega_\alpha\\
 \dot{\Pib} &=& -\left(1+\frac{eB\Omega_{\alpha}}{\hbar}\right)\frac{\partial \hat{V}}{\partial \bR} - e\frac{\partial \hat{H}_0}{\partial \Pib}\times \bB,
 \end{eqnarray}
which coincide indeed with the semiclassical equations of motion to linear order in $B$ and $\Omega$ \cite{foot2}.

To complete the quantum description at this order of the expansion, we also need to expand the Hamiltonian (\ref{eq:ham}) to the same order,
\begin{equation}\label{eq:ham2}
 \hat{H}=\hat{H}_0(\bp) + \hat{V}(\br) +\frac{1}{2\hbar}\frac{\partial \hat{V}}{\partial \br}\cdot \left(\Omega_\alpha\times \bp\right) +
 \frac{e}{2\hbar}\frac{\partial \hat{H}_0}{\partial \bp}\cdot\left(\bB\times\br\right),
\end{equation}
which is now expressed in terms of the canonical variables $\br$ and $\bp$ and thus amenable to the usual quantum-mechanical treatment. 
This result coincides with the Hamiltonian obtained by Zhou et al. for $B=0$ \cite{Zhou2015}. 

\subsection{Corrective Berry-curvature terms in the excitonic Hamiltonian }\label{sec:2.2}

The above discussion has direct consequences for the quantum-mechanical description of excitons in multiband systems, where the third term in the Hamiltonian (\ref{eq:ham2}) arises from the mutual interaction between the electron and the hole constituting the (neutral) exciton. The exciton consists of an electron in the valence band, described by the (band) Hamiltonian in the continuum 
\begin{equation}
 \hat{H}_0^e=\frac{\Delta}{2} + \frac{\bp_1^2}{2m_e}
\end{equation}
and a hole in the valence band with
\begin{equation}
 \hat{H}_0^h=\frac{\Delta}{2} + \frac{\bp_2^2}{2m_h},
\end{equation}
where $m_e$ and $m_h$ are the band masses for the electron and the hole, respectively, and $\Delta$ is the direct gap between the valence and the conduction band. 
The momenta $\bp_1$ and $\bp_2$ are measured from the reciprocal-space position of the conduction-band minimum, which coincides with the valence-band maximum in the present
case of a direct-gap semiconductor. We consider, here, intra-valley excitons, i.e. where both the electron and the hole reside in the same valley. The electron
and the hole interact via an attractive interaction potential $\hat{V}(|\br_1-\br_2|)$ that we consider as isotropic -- one may think of the usual Coulomb interaction with possible
corrections due to screening. In most approaches, one simply considers the Hamiltonian $\hat{H}_0^e+\hat{H}_0^h+\hat{V}$, which constitutes the (2D) hydrogen problem in the case of a pure
Coulomb interaction. However, the interaction induces corrective terms via the Berry curvature,
\begin{equation}\label{eq:HB}
 \hat{H}_B=\frac{1}{2\hbar}\frac{\partial \hat{V}(\rho)}{\partial \rhob} \cdot \left[\Omega_e(\bp_1)\times\bp_1 + \Omega_h(\bp_2)\times\bp_2\right],
\end{equation}
where $\rhob=\br_1-\br_2$ is the relative coordinate and $\rho=|\rhob|$. 

One notices from this equation that relative and center-of-mass motion are not decoupled. Indeed, if we use the usual change in coordinates 
\begin{eqnarray}
 M \bR=m_e \br_1 + m_h\br_2, && \qquad \rhob=\br_1-\br_2 \\
\nonumber
 \bP=\bp_1 + \bp_2, && \qquad \bp/\mu=\bp_1/m_e - \bp_2/m_h,
\end{eqnarray}
in terms of the total mass $M=m_e+m_h$ and the relative mass $\mu=m_em_h/M$, the corrective term (\ref{eq:HB}) becomes
\begin{eqnarray}
\nonumber
 \hat{H}_B &=& \frac{1}{2\hbar}\frac{\partial \hat{V}(\rho)}{ \partial \rhob} \cdot \left[\Omega_e\left(\bp + \frac{m_e}{M}\bP\right)\times \left(\bp + \frac{m_e}{M}\bP\right)\right.\\
 &&\left. - \Omega_e\left(-\bp + \frac{m_h}{M}\bP\right)\times \left(-\bp + \frac{m_h}{M}\bP\right)\right],
\end{eqnarray}
where we have already used $\Omega_h(\bq)=-\Omega_e(\bq)$, valid for two-band models. This is in line with the case of massive Dirac fermions -- a particular case of the 
two-band models discussed here --
because a change in the frame of reference for the center-of-mass motion also affects, due to the associated Lorentz contraction, the relative coordinates. While we consider 
excitons in the center-of-mass frame of reference (i.e. $\bP=0$), we briefly comment on the case $m_e=m_h$ relevant in 2D transition-metal dichalcogenides (TMDC). 
One then finds, to lowest order in $\Omega_e$ (i.e. in the absence of gradient terms $\partial\Omega_e/\partial \bp$)
\begin{equation}\label{eq:Hberry}
 \hat{H}_B\simeq \frac{1}{2\hbar}\frac{\partial \hat{V}(\rho)}{ \partial \rhob} \cdot \left[\Omega(\bP/2)\times \bp\right],
\end{equation}
where $\Omega(\bp)=2\Omega_e(\bp)$ can be considered as the exciton Berry curvature. Also here one notices the coupling between center-of-mass and relative coordinates. 
Adding now the Darwin-type term $\Omega(\bq)\nabla_{\rho}^2\hat{V}(\rho)/4$, one obtains, 
in the center-of-mass frame of reference, which we consider from now on, the semiclassical exciton Hamiltonian \cite{Zhou2015,Trushin2018}
\begin{eqnarray}\label{eq:hamB}
 \hat{H}_{X} &=&  \frac{\bp}{2\mu} + \hat{V}(\rho) \\
 \nonumber
 &&+ \frac{1}{2\hbar}\frac{\partial \hat{V}(\rho)}{\partial \rhob} \cdot \left[\Omega(\bp)\times \bp\right]
 + \frac{1}{4}\left|\Omega(\bq)\right|\nabla_{\rho}^2\hat{V}(\rho),
\end{eqnarray}
in terms of the exciton Berry curvature \cite{Yao2008,Garate2011}
\begin{equation}
 \Omega(\bp)=\Omega_e(\bp)+ \Omega_e(-\bp)\simeq 2\Omega_e(0)\simeq \frac{\hbar^2}{\mu \Delta} \be_z, 
\end{equation}
where $\be_z$ is the unit vector in the $z$-direction. Notice that Hamiltonian (\ref{eq:hamB}) describes the binding energy of the exciton, i.e. we have omitted the 
gap energy $\Delta$, which appears in the energy to create a free electron and a free hole. 
From the exciton Hamiltonian (\ref{eq:hamB}), one can already draw some conclusions. Most importantly, the last two terms, which depend on the Berry curvature, 
modify the exciton spectrum as compared to the usual 2D hydrogen model. The latter is retrieved in the limit of large gaps for which the Berry-curvature terms 
vanish as $\Omega \sim 1/\Delta$, i.e. when the gap becomes by far the largest energy scale. Furthermore, these corrective terms do not play a role in direct-gap
semiconductors, where the gap is situated at a time-reversal-invariant momentum, in which case the Berry curvature vanishes. Finally, the Berry-curvature terms 
play a minor role in states with large angular momentum where the average distance between the electron and the hole increases. This can easily be seen for a 
$1/r$-Coulomb-type potential where the last two terms of (\ref{eq:hamB}) scale as $1/r^3$, while the direct Coulomb term scales as $1/r$ and the centrifugal terms in
the usual manner as $1/r^2$ \cite{Trushin2018,Trushin2016}. Notice also that the chirality of the second last term breaks the rotational symmetry, such that the $m\leftrightarrow -m$
degeneracy is lifted \cite{srivastava2015}.

Notice that, in the above discussion, we have not explicitly taken into account the spin or the valley of the electron and hole constituting the exciton. If we consider 2D 
TMDC, the strong spin-orbit coupling leads to a locking between spin and valley. This means that, if we consider e.g. spin-up electrons in the 
$K$ valley (i.e. the highest valence band is occupied by spin-up particles), the spin is reversed when one wants to treat excitons in the $K'$-valley. The above discussion then 
remains valid, but we need to change the sign of the Berry curvature, such that the last term in Eq. (\ref{eq:hamB}) acquires a minus sign. 

\subsection{Revisiting neutral excitons in monolayer WS$_2$}

Before generalizing the above description for neutral excitons to trions, let us first revisit, in view of the Berry-curvature corrections, 
the exciton binding energy in monolayer WS$_{2}$ \cite{hichri}.
We start with some remarks on the pure Coulomb potential $\hat{V}(\rho)=-e^2/\kappa \rho$, which can to large extent be treated analytically. Here, the 
effective dielectric constant $\kappa=(\varepsilon_{sub}+\varepsilon_{vac})/2$ is the average of the the dielectric permittivity of 
the substrate, $\varepsilon_{sub},$ and that of the vacuum, $\varepsilon_{vac}$. In the absence of Berry-curvature corrections, one obtains thus the usual 2D
hydrogenic spectrum $E_{n,m}=-{\rm Ry}/(n+1/2)^2$, which is degenerate in the angular-momentum quantum number $m$ and only depends on the principal quantum number $n$. 
Here, ${\rm Ry}\simeq 13 ~{\rm eV}\times (\mu/m_0)/ \kappa^{2}$ is the effective Rydberg 
energy for a reduced mass of $\mu=0.17 m_0$ and  $\kappa \simeq 1.55$, relevant for WS$_2$ \cite{Trushin2018, Chernikov2014}. 
The degeneracy in the angular momentum $m$ is lifted due to the chiral Berry-curvature term and splits the level into 
$\Delta_{n,m}=|E_{n,m+} - E_{n,m-}|$ according to the handedness of the angular momentum \cite{Zhou2015,srivastava2015}. 
Using the atomic orbital terminology $s$ (for $m=0$), $p_{\pm}$ (for $m=\pm 1$), $d_{\pm}$ (for $m=\pm 2$) ..., we find $\Delta_{2p}=(64/81)(\Omega/a_{B}^{2}){\rm Ry}$, in 
agreement with Ref. \cite{Zhou2015} and $\Delta_{3p}=(64/375)(\Omega/a_{B}^{2}){\rm Ry}$. The numerical values can be found in the second line of Tab. \ref{tab1}, for a value
of $a_B=0.5{\rm \AA}\times (m_0/\mu)\kappa\simeq 5$ \AA\  and an upper bound of $\Omega =10$ \AA$^2$ for the Berry curvature \cite{foot3}.

\begin{table*}[htb]
\caption {\label{tab1} The effect of Berry curvature on the energy splitting of $np$ states.}
\begin{tabular}{ccccc}
\hline
\hline
             $\Delta_{n,p}$ (in meV)    &\textit{n}=2• • & \textit{n}=3• &\textit{n}=4• & \textit{n}=5•  \\
\hline
       
 with locally-screened  interaction •     &321 •        & 69•        & 25 •         & 12 \\ 
 with non locally-screened  interaction • &17 • & 5• & 1.8• & 1.6\\     
 
\hline
\hline

\end{tabular}
\end{table*} 

The above numerical values become smaller when non-local screening effects are taken into account, which furthermore lift the degeneracy in the quantum number $m$ already in 
the absence of Berry-curvature corrections, while keeping the degeneracy in $m\leftrightarrow -m$. 
The appropriate electron-hole interaction $\widehat{V}(\rho)$ is given by the Keldysh potential \cite{Keldysh}
\begin{eqnarray}\label{eq:V}\nonumber
\widehat{V}(\rho) &=& -\frac{e^{2}}{\kappa2\pi}\int\frac{e^{iq\cdot\rho}d^{2}q}{q(1+q \lambda_{s})} \\
&=& -\frac{\pi e^{2} }{2 \kappa \lambda_{s}}\left[ H_{0}(\rho/ \lambda_{s})-Y_{0}(\rho/ \lambda_{s})\right]
\end{eqnarray}
where $H_{0}(x)$ and $Y_{0}(x)$ are the Struve and Bessel function of the second kind, respectively, and 
$ \lambda_s$, which can be related to the 2D polarizability of the monolayer material, gives a crossover length scale between the long and short range Coulomb interaction. 
The expressions
$\nabla _{\rho}\widehat{V}(\rho)$ and $\nabla _{\rho}^{2}\widehat{V}(\rho)$ in the Berry and 
Darwin terms can be directly calculated and give 
\begin{multline}
\frac{•\partial \widehat{V}(\rho)}{•\partial\rho}=-\frac{e^{2}}{•\kappa \lambda_{s}}\int\frac{•dq}{1+q \lambda_{s}•}\frac{\partial J_{0}(q \rho)}{•\partial\rho}=\frac{e^{2}}{•\kappa \lambda_{s}}
\int\frac{•q\,dq}{1+q \lambda_{s}•}J_{1}(q\rho)\\
\frac{•\partial^{2} \widehat{V}(\rho)}{•\partial\rho^{2}}=\frac{e^{2}}{•\kappa \lambda_{s}}\int\frac{•q\,dq}{1+q \lambda_{s}•}\left[ qJ_{0}(q\rho)-\frac{J_{1}(q\rho)}{•\rho}\right] 
\end{multline}
where $J_{\nu}(X)$ are Bessel functions of the first kind. The asymptotic behavior of $\widehat{•V}(\rho)$ is  $\widehat{V}(\rho\rightarrow\infty)\sim 1/\rho$ 
and $V(\rho\rightarrow 0)\sim-\frac{1}{•\lambda_{s}}[ \ln ( \rho /2 \lambda_{s} ) +\gamma] $ where $\gamma\sim0.5772$ is Euler's constant. 
The crossover between these two behaviors is characterized by the length scale $\lambda_s $. By the simplest possible matching of the two asymptotic behaviors and 
to avoid the divergence of the integral of the Bessel function $qJ_{\nu}(q\rho)$, we can construct an approximated expression for $V(\rho)$ in terms of elementary functions  \cite{Cudazzo2011} 
\begin{equation}\label{eq:Vapprox}
•\widehat{V}(\rho)=\frac{e^{2}}{\kappa•\lambda_{s}}\left[ \ln \left( \frac{\rho }{\rho+\lambda_s•}\right) +(\gamma-\ln  2)e^{-\rho /\lambda_{s}}\right]
\end{equation}
which gives an accurate description of the screening potential also at intermediate values of  $(\rho /\lambda_{s})$. In this form, the resulting energy of excitonic systems is a function of only $\lambda_s$ and $\kappa$. Now, to determine the new eigenvalues of the Hamiltonian given in Eq. (\ref{eq:hamB}), one needs calculate the matrix 
elements of $\widehat{H}_{X}$ in the basis of 2D hydrogenic eigenfuction, which become the exact ground state wavefunction in the limit of weak screening. 
The Berry term $\widehat{H}_{B}$ [Eq. (\ref{eq:Hberry})], has only off-diagonal matrix elements which are non-zero; it acts on exciton states with opposite angular momentum and thus
lifts the degeneracy between $np_{+}(nd_{+})$ and $np_{-}(nd_{-})$ states, This can be seen in the following equation for the Berry-curvature correction  
\begin{equation}
\frac{\widehat{H}_{B}}{\rm Ry}\simeq \frac{i  a_{B}\Omega}{•\lambda_{s}}\left(  \frac{1}{•\rho}- \frac{1 }{\rho+\lambda_s•} -\frac{1}{•\lambda_{s}}(\gamma-\ln 2)e^{-\rho/ \lambda_{s}}\right)  
\frac{1}{•\rho}\dfrac{\partial}{\partial\theta•}
\end{equation}
In contrast to the hydrogenic model with local conventional Coulomb interaction, the energy splitting 
calculated with the Keldysh potential is substantially smaller, for $ \lambda_{s}=28$\AA,  $\Delta_{2p}=\vert E_{2p_{+}}-E_{2p_{-}}\vert=17$ meV  
and almost vanishes for  $n\geqslant4$ (see Tab. \ref{tab1}).  
Consequently, one can consider the Berry curvature correction using  Keldysh interaction as just a perturbation, which only introduces a small splitting between the $m\neq0$ states, 
in comparison with that using the conventional 2D hydrogenic  potential. The obtained $2p$ splitting is consistent with the reported splitting of 10 meV  (15 meV) 
in MoS$_{2}$ (WS$_{2}$) \cite{Malic2017}. 
The Darwin term  is proportional to $\nabla _{\rho}^{2}\widehat{V}$ and leads to an energy shift depending on the quantum number $n$ \cite{Zhou2015}.

\begin{figure}[!h]
  \includegraphics[width=0.5\textwidth]{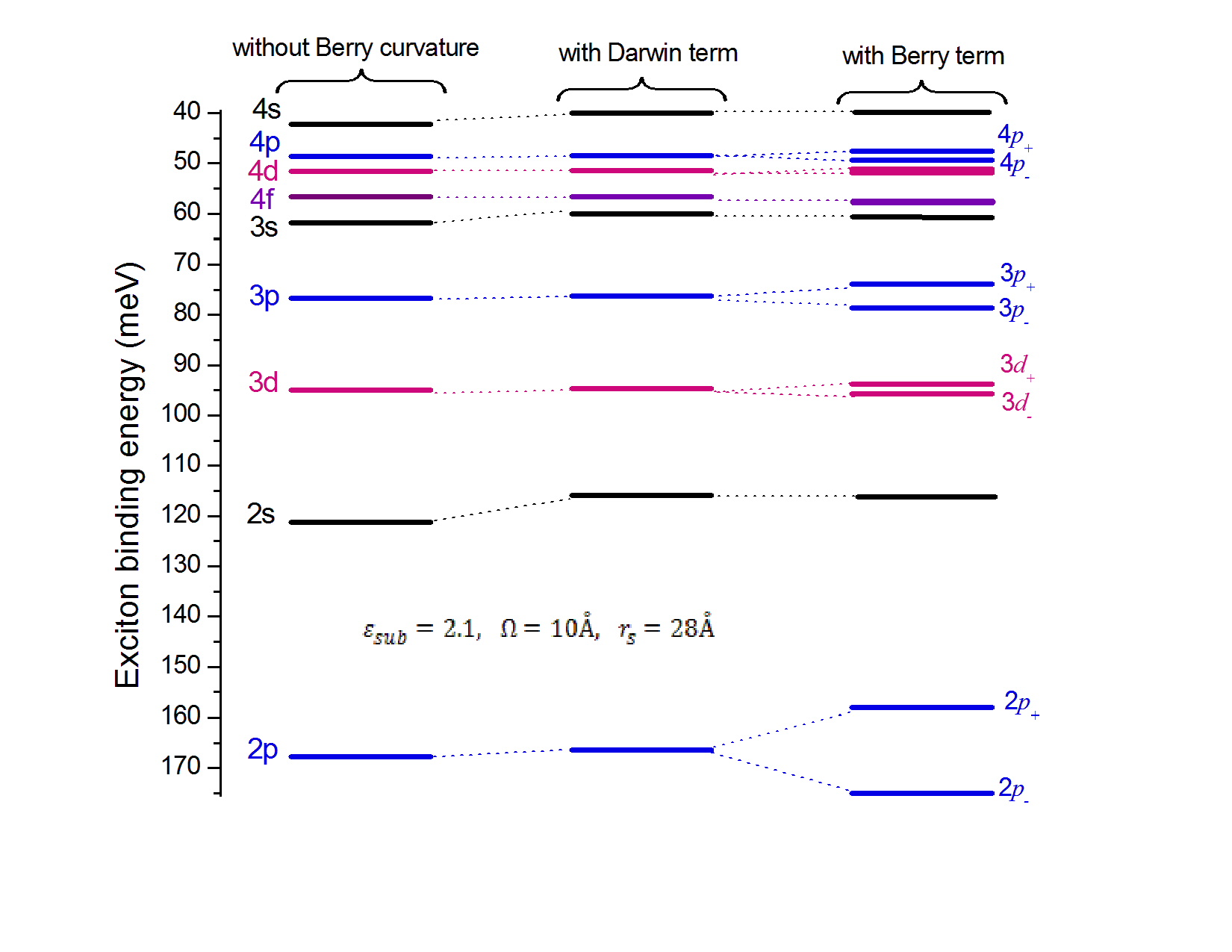}
\caption {Theoretical excitonic spectrum reported for WS$_{2}$ monolayer deposed on the SiO$_{2}$ substrate $\varepsilon_{sub}=2.1$ and exposed to the air  
$\varepsilon_{vac}=1$ for the screening length  $\lambda_{s}=28 $ \AA\ and $a_{B}=5$ \AA. The results are obtained by numerical diagonalization 
of the Hamiltonian given in Eq. (\ref{eq:hamB}), using the non-local screening potential and taking into account the Berry curvature correction. We show a mixing and splitting 
of $np$ and $nd$ states due to the Berry curvature.}
\label{fig:1}       
\end{figure}

In Fig. \ref{fig:1}, we plot the calculated first positions of the exciton binding energy $B_{nm}=-E_{nm}$ with various principal and 
orbital angular momentum quantum numbers, $n$ and $m$, respectively. Here, $E_{nm}$ are obtained by the diagonalization of the Hamiltonian (\ref{eq:hamB}). 
The large binding energy in monolayer TMDCs results from the enhanced Coulomb interaction due to the strong quantum confinement, reduced dielectric screening and 
heavy effective masses \cite{Stier2016}. This reduction of the exciton binding energy is less pronounced 
for higher energy exciton states $(B_{2s}<B_{1s})$. We recall that, while the 
usual $1/\rho$ Coulomb potential gives equal energies $E_{n,m}=E_{n}$, this $(2n-1)$-fold degeneracy is lifted by the screened potential [Eqs. (\ref{eq:V}) and (\ref{eq:Vapprox})] 
which originates from the weak dielectric broadcast in the 2D limit \cite{Keldysh}. Our calculations show that the energy 
of the $m = 1$ ($m=2$) excited state is lower than that of the $m = 0$ ($m = 1$) excited state i.e. $E_{2p}<E_{2s} (E_{3d}<E_{3p}...)$. 
The degeneracy between $2s$ and $2p$ states is lifted with  $2p_{-}$ states lying about 37 meV below the $2s$ states. This result is in agreement with other 
theoretical investigations where $E_{2s}-E_{2p}=35$meV \cite{srivastava2015, Malic2017,Xiao2017, Trushin2016}.

Upon the inclusion of the non-local potential and Berry curvature, states with opposite angular momentum mix and split much like in the case of an orbital Zeeman effect. 
The Darwin term acts mainly on the $s$-states, due to its very local action \cite{Trushin2018}. Principally, the energy position of $1s$ exciton state shows a decreasing
shift in response to this extra term which, would otherwise remain unaffected by the Berry term. However, the $1s$ ground state  $ \xi_{1s}=\Delta-B_{1s} $ shifts 
from 2.079 to 2.101 eV. 
Consequently, the $1s$ exciton binding energy decreases by 22meV as compared to that obtained in our previous work \cite{hichri}, where the Berry and Darwin terms were neglected. 
This results in binding energies around 290 meV. For a better comparison with experimental measurements and other theoretical findings, Tab. \ref{tab2} summarizes
the exciton binding energies in monolayer WS$_{2}$ and WSe$_{2}$ for a wide range of screening lengths. Varying the screening length and including the Berry correction 
dramatically changes the results. A general trend appearing in our calculations is that when the screening length exceeds the Bohr radius the exciton binding energy is weak, 
while for smaller screening lengths ($\lambda_s\sim  a_{B}$) the exciton becomes more strongly bound. 
For a suspended WX$ _{2} $ monolayer, the exciton binding energy reaches 900meV even if the Berry correction is taken into account.  
On the other hand, the Berry correction decreases significantly with increasing 
exciton binding energy, i.e. for small $\lambda_s$.  However, depending on the screening length, the adjustement with Berry correction cannot be made because in a certain 
range the exciton binding energy differs significantly from the observed value and thus does not agree with experiment. These values can serve as a benchmark for theory.

\begin{table*}[htb]
\caption {\label{tab2} Exciton binding energies (given in meV) for both WS$_{2}$ and WSe$_{2}$ monolayer deposed on SiO$ _{2} $ substrate. The effective screening radius $ \lambda_{s} $ varies between $ \sim5 $ and $ \sim25 $\AA. Experimental and theoretical results from literature are collected for comparison.
The reduced mass used for WSe$ _{2} $ monolayer are listed in Ref. \cite{Hanan2018}.}
 \begin{center}
\begin{tabular}{ccccc}
\hline
\hline
              & \multicolumn{2}{c}{WS$_{2}$} & \multicolumn{2}{c}{ WSe$_{2}$ }   \\
\hline
         & $\Omega=0$ & $\Omega=10{\rm \AA}^{2}$   & $\Omega=0$ & $\Omega=10{\rm \AA}^{2}$ \\
 $\lambda_{s}=a_{B}$       &626   & 480  & 840 & 607   \\
 $\lambda_{s}=1.5a_{B}$    &539   & 445  & 724 & 564   \\      
 $\lambda_{s}=2.5a_{B}$    &439 • & 386• & 590 & 493   \\
 $\lambda_{s}=3.5a_{B}$    &376 • & 341• & 505 & 437   \\
 $\lambda_{s}=4.a_{B}$     &352•  & 322• & 473 & 414   \\
 $\lambda_{s}=5.a_{B}$     &313•  & 290• & 420 & 375   \\
 Experiment              & \multicolumn{2}{c}{320 \cite{Chernikov2014} -700 \cite{Ye2014,Hanbicki2015}} & \multicolumn{2}{c}{ 370 \cite{He2014}, 600 \cite{Wang2014}}   \\
 Theory              & \multicolumn{2}{c}{509 \cite{Kyl2015}, 523 \cite{Zhang2015},830 \cite{Hanbicki2015},1050 \cite{Ramasubramanium2012}} & \multicolumn{2}{c}{ • 456 \cite{Kyl2015}, 470 \cite{Zhang2015}, 790 \cite{Hanbicki2015}}   \\ 
\hline
\hline

\end{tabular}
\end{center}
\end{table*}

\section{Trion Hamiltonian in the continuum limit} 
\label{sec:trions}

In the presence of residual free charge carriers, excitons interact with the surrounding charges and can form charged excitons \cite{Singh2016, Stebe}. The strong Coulomb interaction between the electron and hole leads to a larger trion binding energy. We consider here negatively charged trions that 
consist of two electrons and one hole all of which reside in the same valley. The approach is easily generalized to positively charged excitons. 
In this case, we have two electron contributions  
\begin{equation}
•\widehat{H}_{0}^{ej}=\frac{\Delta}{2} + \frac{\textbf{•p}_{ej}^{2}}{•2m_{e}}
\end{equation}
for the electrons $e_{1}$ and $e_{2}$, i.e. \textit{j} = 1, 2, in the same band with band mass $m_{e}$ and a contribution  
\begin{equation}
•\widehat{H}_{0}^{h}=\frac{\Delta}{2} + \frac{\textbf{•p}_{h}^{2}}{•2m_{h}}
\end{equation}
from the hole in the valence band. The interaction Hamiltonian contains now three terms
\begin{equation}
•\widehat{V}(\rho)=\widehat{V}(\rho_{1})+\widehat{V}(\rho_{2})-\widehat{V}(\vert \boldsymbol{\rho}_{1}-\boldsymbol{\rho}_{2}\vert)
\end{equation}
where $\boldsymbol{\rho}_{1}=\textbf{r}_1-\textbf{r}_h$ and $\boldsymbol{\rho}_{2}=\textbf{r}_2-\textbf{r}_h$ are the relative coordinates between the two electron positions 
$\textbf{r}_1$, $\textbf{r}_2$ and the hole position $\textbf{r}_h$,
\begin{equation}\label{eq:relCoor}
•\boldsymbol{\rho}_{j}=\textbf{r}_j-\textbf{r}_h, \qquad \boldsymbol{\rho}_{1}-\boldsymbol{\rho}_{2}=\textbf{r}_1-\textbf{r}_2
\end{equation}
Without the Berry-curvature correction, this 
trion Hamiltonian can be brought into the form  
\begin{multline}\label{eq:HT}
•\widehat{H}_{T}= 
\frac{\textbf{•P}^{2}}{•2M_{T}}+\dfrac{\textbf{•p}_{1}^{2}}{2\mu•}+\dfrac{\textbf{•p}_{2}^{2}}{2\mu•}+\dfrac{\textbf{•p}_{1}\cdot\textbf{•p}_{2}}{2 m_h}
\\
\\+ \widehat{V}(\rho_{1})+\widehat{V}(\rho_{2})-\widehat{V}(\vert \boldsymbol{\rho}_{1}-\boldsymbol{\rho}_{2}\vert)+\widehat{H}_{B}^{T}+\widehat{H}_{D}^{T}
\end{multline}
where the term $\widehat{H}_{B}^{T}$ accounts for the trion Berry-curvature corrections, while $\widehat{H}_{D}^{T}$ represents a trion Darwin term that 
arises, in the same manner as for neutral excitons, within relativistic quantum mechanics, but which is beyond the semi-classical description limited to first-order gradient terms in the potential. 
Again, we have omitted the constant term $3\Delta/2$, which represents the energy to create three non-interacting particles because we are interested in the binding energies, i.e. the negative eigenvalues of the 
trion Hamiltonian (\ref{eq:HT}).

The Hamiltonian (\ref{eq:HT}) is expressed in terms of the 
relative coordinates (\ref{eq:relCoor}) and the trion center-of-mass coordinate $\textbf{R}_{T}=\left[ m_{e}(\textbf{r}_{1}+\textbf{r}_{2})+m_{h}\textbf{r}_{h}\right]/M_{T} $, 
where $M_{T}=2m_{e}+m_{h}$ is the total mass. The momenta
\begin{multline}\label{eq:momenta}
\textbf{P}_{T}=\textbf{p}_{e1}+\textbf{p}_{e2}+\textbf{p}_{h}\\
\textbf{p}_{1}=\mu\left( \dfrac{\textbf{p}_{e1}}{m_{e}•}-\dfrac{\textbf{p}_{h}}{m_{h}•}\right),\qquad
 \textbf{p}_{2}=\mu\left( \dfrac{\textbf{p}_{e2}}{m_{e}•}-\dfrac{\textbf{p}_{h}}{m_{h}•}\right)
\end{multline}
are conjugate to \textit{\textbf{R}}, $\boldsymbol{\rho}_{1}$ and $\boldsymbol{\rho}_{2}$, respectively. 


\subsection{Berry-curvature correction}\label{sec:3.1}

Let us now discuss in detail the part $\widehat{H}_{B}^{T}$ in the Hamiltonian, i.e. take into account the Berry curvature. For one of the trion components, this term arises from the interaction potential 
generated by the other two components, and the Hamiltonian thus consists of six contributions,
\begin{eqnarray}\nonumber
•\widehat{H}_{B}^{T} &=& \frac{1}{•2\hslash}\nabla_{\rho_{1}}\widehat{V}\cdot\left[ \Omega_{e}(\textbf{p}_{e1})\times \textbf{p}_{e1} +\Omega_{h}(\textbf{p}_{h})\times \textbf{p}_{h}\right]\\
\nonumber
&& +\frac{1}{•2\hslash}\nabla_{\rho_{1}}{\widehat{V}}\cdot \left[ \Omega_{e}(\textbf{p}_{e2})\times \textbf{p}_{e2} +\Omega_{h}(\textbf{p}_{h})\times \textbf{p}_{h}\right]\\
\nonumber
&& -\frac{1}{•2\hslash}\nabla_{\rho_{3}}\widehat{•V}\cdot\left[ \Omega_{e}(\textbf{p}_{e1})\times \textbf{p}_{e1} +\Omega_{e}(\textbf{p}_{e2})\times \textbf{p}_{e2}\right],\\
 \end{eqnarray}
where we have defined $\boldsymbol{\rho}_{3}=\textbf{r}_1-\textbf{r}_2$ as the relative coordinate between the two electrons \cite{foot4}. Remember that we have opposite Berry curvatures in the 
different bands, $\Omega_{e}(\textbf{k})=-\Omega_{h}(\textbf{k})$, at each wave vector, and we furthermore approximate the Berry curvature by its value at the gap $\Omega_{e}(\textbf{k})\simeq\Omega_{e}(0)$ \cite{foot5}.
The Berry-curvature contribution thus becomes 
\begin{eqnarray}\nonumber
•\widehat{H}_{B}^{T} &\simeq& \frac{1}{•2\hslash}\nabla_{\rho_{1}}\widehat{V}\cdot\left[\Omega_{e}(0)\times\left(\textbf{p}_{e1}-\textbf{p}_{h}\right)\right]\\
\nonumber
&& +\frac{1}{•2\hslash}\nabla_{\rho_{2}}V\cdot \left[\Omega_{e}(0)\times\left(\textbf{p}_{e2}-\textbf{p}_{h}\right)\right]\\
&&-\frac{1}{•2\hslash}\nabla_{\rho_{3}}V\cdot\left[\Omega_{e}(0)\times\left(\textbf{p}_{e1}+\textbf{p}_{e2}\right)\right].
\end{eqnarray}
These expressions turn out to be more complicated than those of the neutral exciton if we aim at expressing them in terms of the momenta (\ref{eq:momenta}), and we therefore use immediately two simplifications, 
valid in the case of 2D TMDC. The first one is to consider conduction and valence bands with the same curvature or band mass, i.e. $m_{e}=m_{h}$; and the second approximation consists of a calculation in the 
center-of-mass frame of the trion, i.e. $\textbf{P}_{T}=0$. We then find
\begin{equation}
\textbf{p}_{e1}+\textbf{p}_{e2}=-\textbf{p}_{h}
\end{equation}
and, with $\mu=m_{e}/2$,
\begin{equation}
\textbf{p}_{h}=-\frac{2}{•3}\left( \textbf{p}_{1}+\textbf{p}_{2}\right) .
\end{equation}
We can thus express the Berry-curvature contribution in terms of the excitonic Berry curvature $\Omega(0)=2\Omega_{e}(0)$, as in the case of the neutral exciton, as \cite{foot6}
\begin{eqnarray}
\nonumber
•\widehat{H}_{B}^{T} &\simeq& \frac{1}{•2\hslash} \nabla_{\rho_{1}}\widehat{V}\cdot\left[\Omega(0)\times\textbf{p}_{1}\right] +\frac{1}{•2\hslash} \nabla_{\rho_{2}}\widehat{V}\cdot \left[\Omega(0)\times\textbf{p}_{2}\right] \\
&& -\frac{1}{•2\hslash}\nabla_{\rho_{3}}\widehat{V}\cdot \left[\frac{•\Omega(0)}{3}\times\left(\textbf{p}_{1}+\textbf{p}_{2}\right)\right],
\end{eqnarray}
which can be further simplified by getting rid of the redundant variable $\rhob_3=\rhob_1 - \rhob_2$, such that 
\begin{eqnarray}\label{eq:trionBerry}
\nonumber
 \hat{H}_B &\simeq&  - \frac{1}{2\hbar}\left( \nabla_{\rho_1}\hat{V} -\frac{1}{3}\nabla_{\rho_1-\rho_2}\hat{V}
 \right) \cdot \left[ \Omega(0)\times \bp_1\right] \\
 \nonumber
 &&-\frac{1}{2\hbar}\left(\nabla_{\rho_2}\hat{V} -\frac{1}{3}\nabla_{\rho_2-\rho_1}\hat{V} \right) 
 \cdot \left[ \Omega(0)\times \bp_2\right],\\
\end{eqnarray}
where the Berry curvature can be expressed explicitly as
\begin{equation}
\boldsymbol{\Omega}(0)=\zeta \frac{\hslash^{2}}{\mu \Delta}\textbf{e}_{z}=2\zeta \frac{\hslash^{2}}{m_{e} \Delta}\textbf{e}_{z}=2 \boldsymbol{\Omega}_{e}(0).
\end{equation}
The sign $\zeta=\pm$ takes into account whether we consider spin-up particles in the $K$-valley $(\zeta=+)$ or spin-down particles in the $K'$-valley $(\zeta=-)$. 
Equation (\ref{eq:trionBerry}) is the main result of this section. For the pure
Coulomb potential with
\begin{equation}
\widehat{V}(\rho)=-\frac{\textit{e}^{2}}{•\varepsilon\rho}, \qquad  \boldsymbol{\triangledown}_{\rho}\widehat{V}=\frac{\textit{e}^{2}}{•\varepsilon\rho^{2}}\frac{\boldsymbol{\rho}}{\rho•}
\end{equation}
the Berry-curvature term reads
\begin{eqnarray}
\widehat{H}_{B} &=& -\zeta\frac{\hslash e^{2}}{•2\mu \varepsilon\Delta}\left[ \frac{1}{\rho_{1}^{3}•}(\textbf{p}_{1}\times\rho_{1})_{z}+\frac{1}{\rho_{2}^{3}•}(\textbf{p}_{2}\times\rho_{2})_{z}\right]\\ 
\nonumber
&& +\zeta\frac{\hslash e^{2}}{•6\mu \varepsilon\Delta \vert\rho_{1}-\rho_{2}\vert^{3}}\left[ (\textbf{p}_{1}+\textbf{p}_{2})\times(\rho_{1}-\rho_{2})\right]_{z}
\end{eqnarray}
where the subscript \textit{z} indicates the \textit{z}-component of the
vector product.
\begin{figure}[!h]
  \includegraphics[width=0.5\textwidth]{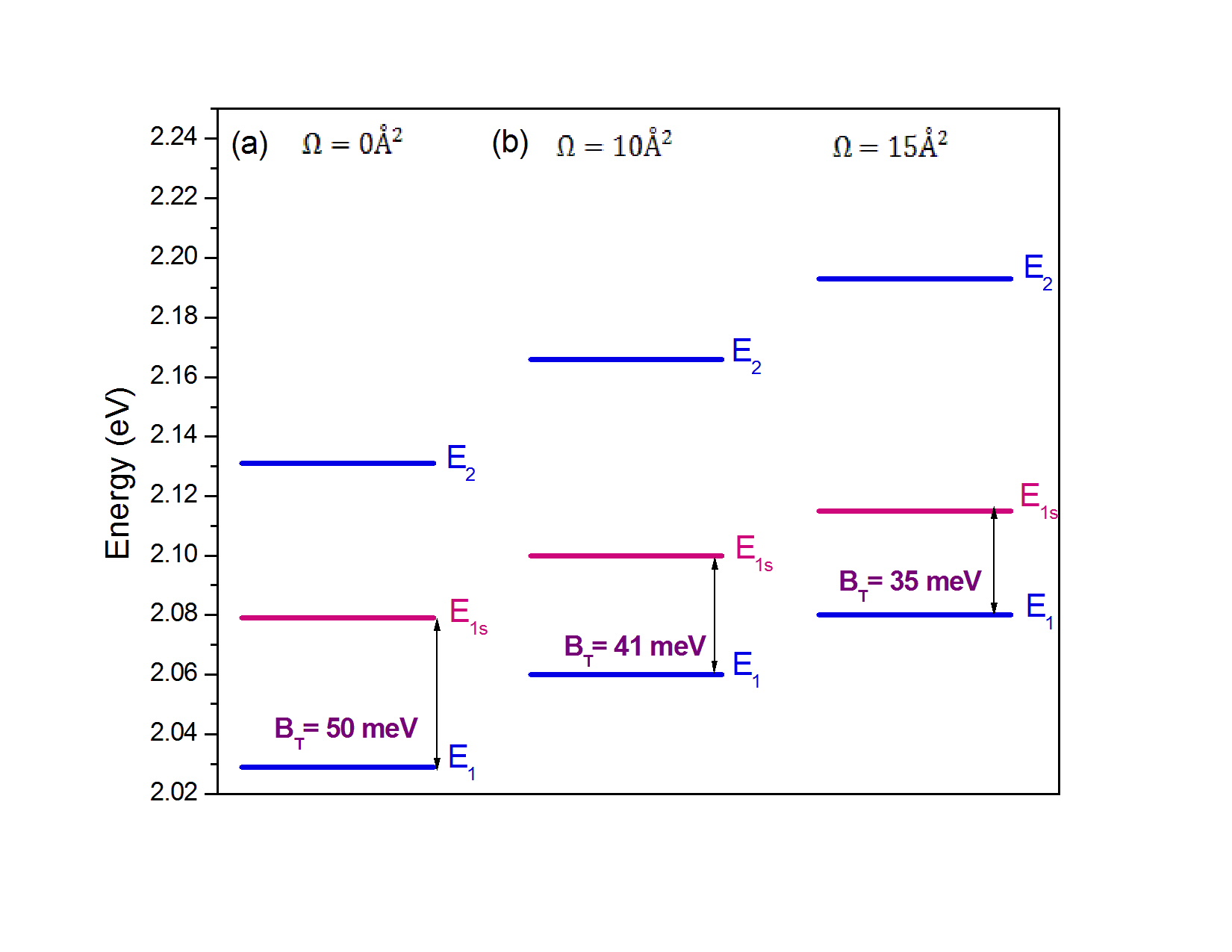}
\caption {Energies of low-lying trion states E$ _{1T} $ and E$ _{2T} $, considering the ground exciton state E$ _{1s} $. Calculations are performed on WS$_{2}$ monolayer using $ \varepsilon_{sub}=2.1, \lambda_{s}=28 $\AA , a) $ \Omega=0 $, b) $ \Omega= $10\AA$ ^{2} $ and $ \Omega= $20\AA$ ^{2} $.}
\label{fig:2a}       
\end{figure}
\begin{figure}[!h]
  \includegraphics[width=0.5\textwidth]{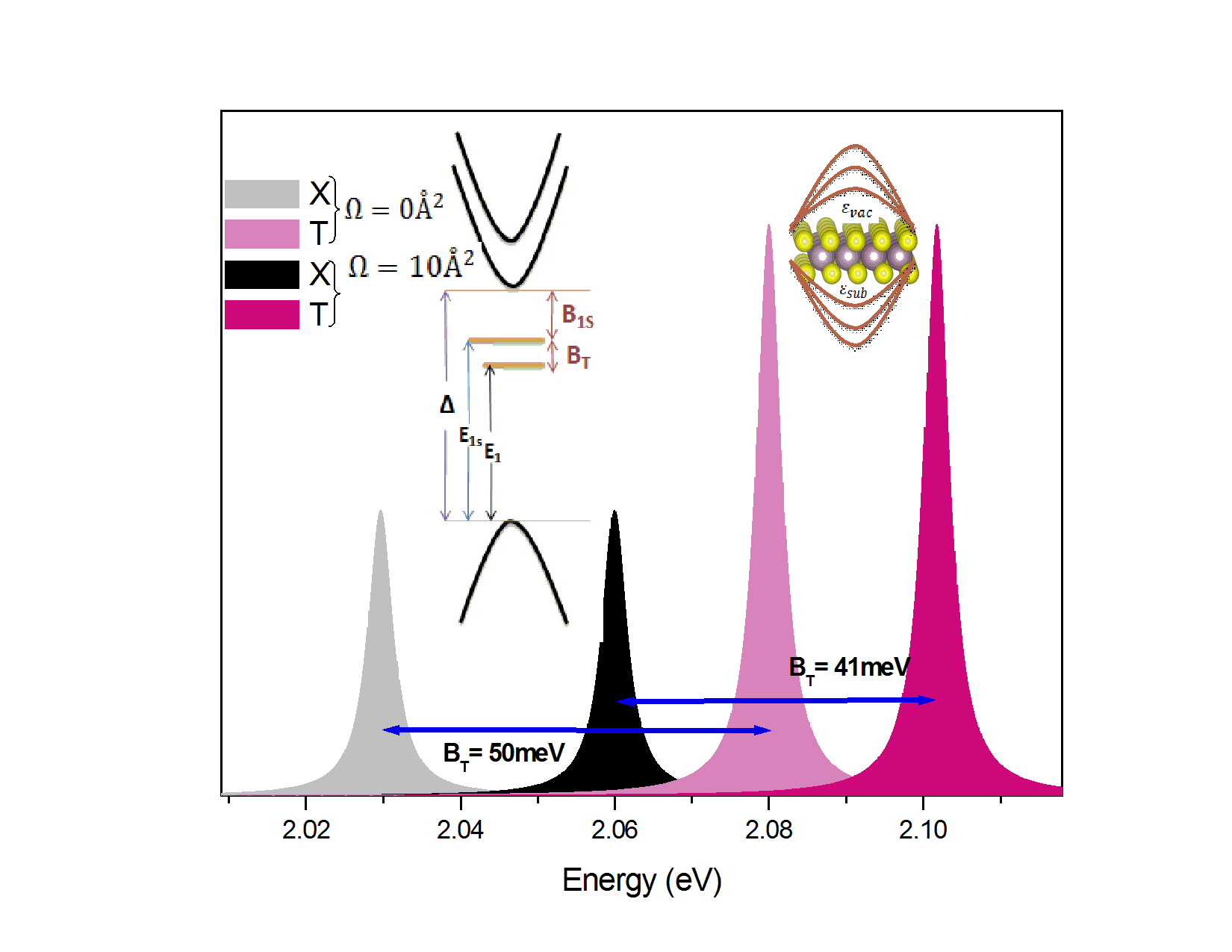}
\caption {Spectrum of a WS$_{2}$ monolayer  showing the effect of Berry correction on exciton X and trion T energies. The parameters used are $ \lambda_{s}=28 $\AA, $ \varepsilon_{sub}=2.1 $ and $ m_{h}=m_{e}=0.34 $.}
\label{fig:2c}       
\end{figure}

\subsection{Darwin term}\label{sec:3.2}

Let us now discuss the role of the Darwin terms, which also arise at linear order in the Berry curvature but as second derivatives of the interaction potential. 
The Darwin term of the \textit{j}-th particle reads $\vert\Omega_{j}\vert\nabla^{2}\widehat{V}(\textbf{r}_{j})/4$, where the potential is again created by the remaining two particles, such that one obtains
\begin{eqnarray}\label{eq:trionDar1}
\nonumber
•H_{D}^{T} &=& \frac{\vert\Omega_{e}(0)\vert}{•4}\left[ \nabla_{\textbf{r}_{1}}^{2}\widehat{V}(\rho_{1})+\nabla_{\textbf{r}_{h}}^{2}V(\rho_{1})\right] \\
&&+\frac{\vert\Omega_{e}(0)\vert}{•4}\left[ \nabla_{\textbf{r}_{2}}^{2}\widehat{V}(\rho_{2})+\nabla_{\textbf{r}_{h}}^{2}V(\rho_{2})\right]\\
\nonumber
&&-\frac{\vert\Omega_{e}(0)\vert}{•4}\left[ \nabla_{\textbf{r}_{1}}^{2}\widehat{V}(\vert\rho_{1}-\rho_{2}\vert)+\nabla_{\textbf{r}_{2}}^{2}\widehat{V}(\vert\rho_{1}-\rho_{2}\vert)\right]
\end{eqnarray}
where we have again used $\Omega_{e}(\textbf{p})=-\Omega_{h}(\textbf{p})\simeq\Omega(0)/2$. Since 
$\nabla_{\textbf{r}_{1}}^{2}\widehat{V}(\rho_{1})=\nabla_{\textbf{r}_{h}}^{2}\widehat{V}(\rho_{1})=\nabla_{\rho_{j}}^{2}\widehat{V}(\rho_{1})$ the Darwin term of the trion Hamiltonian reads \cite{foot6}
\begin{equation}\label{eq:trionDar2}
•\widehat{H}_{D}^{T}=\frac{\vert\Omega(0)\vert}{•4}\left[ \nabla_{\rho_{1}}^{2}\widehat{V}(\rho_{1})+\nabla_{\rho_{2}}^{2}\widehat{V}(\rho_{2})-\nabla_{\rho_{3}}^{2}
\widehat{V}(\rho_{3})\right] 
\end{equation}
Collecting both the Berry-curvature correction and the Darwin term, the trion Hamiltonian is
\begin{eqnarray}
\nonumber
\widehat{H}_{T} &=& \frac{3\Delta}{2}+\frac{\textbf{•p}_{1}^{2}}{2\mu•}+\widehat{V}({\rho_{1}})\\
\nonumber
&&\qquad +\frac{1}{•2\hslash}\nabla_{\rho_{1}}\widehat{V}\cdot\left[ \Omega(0)\times \textbf{p}_{1}\right] +\frac{\vert\Omega(0)\vert}{•4}\nabla_{\rho_{1}}^{2}\widehat{V}(\rho_{1})\\
\nonumber
&& +\frac{\textbf{•p}_{2}^{2}}{2\mu•}+\widehat{V}({\rho_{2}})\\
\nonumber
&&\qquad +\frac{1}{•2\hslash}\nabla_{\rho_{2}}\widehat{V}\cdot\left[ \Omega(0)\times \textbf{p}_{2}\right] +\frac{\vert\Omega(0)\vert}{•4}\nabla_{\rho_{2}}^{2}\widehat{V}(\rho_{2})\\
&&+\frac{\textbf{•p}_{1}\cdot \textbf{•p}_{2}}{2m_{h}•}-\widehat{V}(\vert{\rho_{1}}-{\rho_{2}}\vert)\\
\nonumber
&&~ -\frac{1}{•2\hslash}\nabla_{\rho_{3}}\widehat{V}\cdot\left[ \Omega(0)\times (\textbf{p}_{1}+\textbf{p}_{2}) \right] -\frac{\vert\Omega(0)\vert}{•4}\nabla_{\rho_{3}}^{2}\widehat{V}(\rho_{3})
\end{eqnarray}
This trion Hamiltonian can be easily expressed as a function of the exciton Hamiltonian (\ref{eq:hamB}) 
\begin{eqnarray}\label{eq:HTb}
 •\widehat{H}_{T} &=& -\frac{\Delta}{2} + \sum _{i=1,2}\widehat{H}_{X_{i}}+\frac{\textbf{•p}_{1}\cdot \textbf{•p}_{2}}{2m_{h}•}-\widehat{V}(\vert{\rho_{1}}-{\rho_{2}}\vert)\\
 \nonumber
&&- \frac{1}{•2\hslash}\nabla_{\rho_{3}}\widehat{V}\cdot\left[ \Omega(0)\times (\textbf{p}_{1}+\textbf{p}_{2})\right]
 -\frac{\vert\Omega(0)\vert}{•4}\nabla_{\rho_{3}}^{2}\widehat{V}(\rho_{3})
\end{eqnarray}

 As a first step towards solving the eigenvalue equation of the relative motion, we use a wave function expansion technique; it is factorized into

\begin{eqnarray}\nonumber
\psi_{T}(\rho_{1},\rho_{2}) &=&\sum_{\tilde{n},\tilde{m}} b_{\tilde{n}\tilde{m}•}\dfrac{1}{•\sqrt{2}} \lbrace \chi_{1\tilde{s}}(\rho_{1}) \chi_{\tilde{n},\tilde{m}}(\rho_{2})+ \\
&&\chi_{\tilde{n},\tilde{m}}(\rho_{1})\chi_{1\tilde{s}}(\rho_{2})\rbrace,
\end{eqnarray}
where $\chi_{\tilde{n},\tilde{m}}=\sum_{n,\vert m\vert <n}c_{nm}\varphi _{n,m}$ 
 is the wave function solution of the exciton  Hamiltonian, expanded in terms of 2D-hydrogenic state $\varphi _{n,m}(\rho,\theta)$. 
 The number  $\tilde{n},\tilde{m}$ refers  to the dominant contribution of the coefficients $c_{nm}$ to the excitonic function \cite{hichri}. 
We emphasize that the trion wavefunction is a symmetric combination of only the $n \tilde{s} $ states. We denote the trion as symmetric in accordance 
 with the symmetry of the wave function. According to Courtade et al.  \cite{Courtade2017} the symmetric trion is stable within the effective mass approximation. 
 Moreover, the trion wavefunction with symmetric combination is the ground state. Since, because we disregard the instable antisymmetric trion, which can be represented by 
 the antisymmetric combination of $ m\neq0 $ states, hereafter the Berry correction is limited to the Darwin term contribution. To find the negative trion eigenvalues  $E_{N}$, 
 we have diagonalize a $5\times 5$ Hamiltonian matrix. For a given in-plane trion center-of-mass wavevector $\textbf{P}_{T}= 0$, we calcultate in the following
 the trion binding energies for both WS$_{2}$ and WSe$_{2}$ monolayer.  

\subsection{Calculation of the trion binding energies}\label{sec:3.3}

The trion binding energy $B_T$ is conventionally introduced as the difference between the trion energy $E_1$, i.e., the ground eigenenergy of the Hamiltonian (\ref{eq:HTb}), 
and the energy of the neutral exciton (\ref{eq:hamB}). In other words, the eigenvalue of the trion bound states is $E_1=-(B_{1s}+B_T )$. Our study takes into account 
the strong dielectric contrast between the vacuum on top of the monolayer and the substrate below it, 
leading to a large trion binding energy close to that measured experimentally. We treat below $\lambda_s$ and $\varepsilon_{sub}$ as parameters of the theory, see 
Tab. \ref{tab3} and Tab. \ref{tab4} (Tab. \ref{tab5}) for discussion of particular values of WS$ _{2} $ (WSe$ _{2} $) monolayer. 
As predicted in Eq. (30), the resulting trion binding energy is a function of the screening length, the Berry curvature and probably the electron-hole mass ratio. 
For the WS$ _{2} $ layer deposed on the top of the SiO$ _{2} $ substrate ($ \varepsilon_{sub} $= 2.1) and exposed to the air ($ \varepsilon_{vac} $ = 1), 
the energies of the first states of trions are plotted in Fig. \ref{fig:2a}. The screening length $\lambda_{s}$ is fixed at $ 28$\AA, in the absence of the Darwin correction, 
the trion ground-state energy is $E_{1T}=2.029$ eV and the first excited state is $E_{2T}=2.131 eV$. In the trion spectrum, we show that the exciton energy $ E_{1s} $ 
(around 2.079 eV in this case) is always located far from $E_{2T}$. Therefore, we will focus only on the trion ground-state energy $E_{1T}$. By including the Berry correction, 
a blue-shift from the lowest trion energies is observed following the enhancement of the exciton energy. A monotone increasing of the energy states with Berry curvature 
is explained by the analytical form of Darwin term given by Eq. (\ref{eq:trionDar2}). The effective Darwin correction leads to an energy shift for all the states. 
The form of the lowest trion and exciton spectrum including the Berry curvature correction is shown in Fig. \ref{fig:2c}. The trion binding energy $ B_{T} $ is calculated 
for the same parameters as those used in Fig. \ref{fig:2a}. The state X represents the neutral 1s exciton and the corresponding negatively charged trion labeled as T. 
This plot shows the decrease of $ B_{T} $ by including the Darwin term as well as, the decrease of $ B_{1s} $ by about 22 meV. 
The results are in line with our findings for the neutral excitons, where the inclusion of the Darwin term correction reduces the binding energy. 
Indeed, in the presence (absence) of Berry curvature effects, particularly the Darwin term we obtain 288 meV (310 meV) for the exciton binding energy and 41 meV (50 meV) 
for that of the trion. The parameters used are: $\lambda_s=28$ \AA, $ \varepsilon_{sub}=2.1 $, $ m_{e}=m_{h}=0.34 $ and $ \Omega=10$\AA$ ^{2} $. 
This result converges towards the experimental data given by Molas \textit{et al.} \citep{Molas2017}.

\begin{table*}[htb]
\caption {\label{tab3} Trion binding energies (given in meV) for both WS$_{2}$ and WSe$_{2}$ monolayer. Experimental and theoretical results from literature are collected for comparison.
The reduced mass used are listed in Ref. \cite{hichri,Rasmussen2015} and $\varepsilon_{sub}=2.1$, corresponding to an effective Bohr radius of $ \sim $5\AA . }
 \begin{center}
\begin{tabular}{ccccc}
\hline
\hline
        $\lambda_s (a_{B})$       & \multicolumn{2}{c}{$ \Omega=0 $} & \multicolumn{2}{c}{ $\Omega$=10 \AA$ ^{2} $}   \\
\hline
   system                     & WS$_{2}$      & WSe$_{2}$       &     WS$_{2}$     &   WSe$_{2}$   \\
 $4.5 $ & 67 &  61 & 59  &  54   \\
 $5.0 $  &  61 & 56 & 53  & 49  \\      
 $5.5 $ &  55 &  51 & 46  &  44 \\
 $6.0$  & 49 & 47 & 41  &  40 \\
 $6.5$ & 43 & 41 &  35  & 34  \\
 system & \multicolumn{2}{c}{WS$_{2}$ } & \multicolumn{2}{c}{ WSe$_{2}$  }   \\
 Experiment              & \multicolumn{2}{c}{34 \cite{Zhu2015}-36\cite{Chernikov2014}-45\cite{Zhu2014}-47\citep{Molas2D2017}} & \multicolumn{2}{c}{ 30\cite{Wang2014}-35\cite{Courtade2017}-47\citep{Molas2D2017} }   \\
 Theory              & \multicolumn{2}{c}{31\cite{Falko2017}-34\cite{Zhang2015}} & \multicolumn{2}{c}{ • 26\cite{Courtade2017}-27\cite{Falko2017}-30\cite{Zhang2015}}   \\ 
\hline
\hline

\end{tabular}
\end{center}
\end{table*}

Clearly, the trion binding energies show a large dependence on the Berry curvature, varying by 8 meV or more, in agreement with the experiment 
using SiO$_{2}$ as a substrate \cite{Plechinger2015}.
This trend is generic, as one can see from Tab. \ref{tab3}, where we have given the trion binding energy for various values of the screening length $\lambda_s$. 
In comparison with experimental measurements, our results suggest that the screening length may be around 5.5 a$ _{B}$ and 6.5 a$ _{B}$ for both WS$_2$ and WSe$_2$ monolayer. 
It is important to notice that the main difference between WS$ _{2} $ and WSe$ _{2} $ monolayer resides only on the effective mass of the charge carriers.

\begin{table*}[htb]
\caption {\label{tab4} Exciton and trion binding energies for monolayer WS$_{2}$ exposed to the air and deposed in different substrates described by $\varepsilon_{sub}$ and for 
a range of screening lengths $\lambda_s$.}
 \begin{center}
\begin{tabular}{ccccccccc}
\hline
\hline
       & & &      \multicolumn{2}{c}{$\lambda_s=20$\AA} & \multicolumn{2}{c}{ $\lambda_s=25$\AA } & \multicolumn{2}{c}{ $\lambda_s=30$\AA }  \\
\hline
    $\varepsilon_{sub} (\varepsilon_{0})$ & $ \kappa (\varepsilon_{0})$     & $\Omega$ (\AA$^{2}$)      &     B$_{1s}$(meV)     &   B$_{T}$(meV)   &     B$_{1s}$(meV)     &   B$_{T}$(meV) &     B$_{1s}$(meV)     &   B$_{T}$(meV)    \\
 \hline
   1.5  & 1.25    & 0      &    510    &  88     &     449    &   67   &   402    &  48    \\ 
        &         & 10     &    457    &  70     &     409    &   49   &   371    &  36   \\ 
\hline
  2.1   & 1.55    & 0      &    372    &  72     &     330    &   56   &   297    &  44    \\ 
        &         & 10     &    338    &  61     &     305    &   46   &   278    &  35    \\ 
\hline
  2.5   & 1.75    & 0      &    311    &  62     &     277    &   52   &   250    &  41    \\ 
        &         & 10     &    284    &  55     &     257    &   45   &   235    &  34    \\  
\hline
\hline

\end{tabular}
\end{center}
\end{table*} 

\begin{table*}[htb]
\caption {\label{tab5} Exciton and trion binding energies for monolayer WSe$_{2}$ exposed to the air and deposed in different substrates described by $\varepsilon_{sub}$ 
and for a range of screening lengths $\lambda_s$.}
 \begin{center}
\begin{tabular}{ccccccccc}
\hline
\hline
       & & &      \multicolumn{2}{c}{$\lambda_s=20$\AA} & \multicolumn{2}{c}{ $\lambda_s=25$\AA } & \multicolumn{2}{c}{ $\lambda_s=30$\AA }  \\
\hline
    $\varepsilon_{sub} (\varepsilon_{0})$ & $ \kappa (\varepsilon_{0})$     & $\Omega$ (\AA$^{2}$)      &     B$_{1s}$(meV)     &   B$_{T}$(meV)   &     B$_{1s}$(meV)     &   B$_{T}$(meV) &     B$_{1s}$(meV)     &   B$_{T}$(meV)    \\
 \hline
   1.5  & 1.25    & 0      &    491    &  86     &     433    &   69   &   389    &  52    \\ 
        &         & 10     &    442    &  74     &     397    &   58   &   361    &  42   \\ 
\hline
  2.1   & 1.55    & 0      &    357    &  67     &     318    &   57   &   287    &  47    \\ 
        &         & 10     &    326    &  60     &     295    &   50   &   269    &  40    \\ 
\hline
  4     & 2.5     & 0      &    171    &  26.9   &     155    &   25.7 &   142    &  24.5    \\ 
        &         & 10     &    160    &  26     &     147    &   25   &   136    &  23.9   \\  
\hline
\hline

\end{tabular}
\end{center}
\end{table*} 
Let us now briefly discuss the role of dielectric environment described by the effective dielectric constant $ \kappa $ on the trion and exciton binding energies. 
By taking into account the Berry curvature correction, Tab. \ref{tab4} (Tab. \ref{tab5}) shows different values for the binding energies in WS$_2$ (WSe$_2$) upon 
variation of the average dielectric constant of the surrounding material and secreening lenght. When the screening length vanishes which corresponds to the strictly 
2D limit of a Coulomb problem, the exciton binding energy is $ \sim4 $ Ry and the trion binding energy reaches its maximum values. With increasing $ \lambda_{s} $, 
the Coulomb potential becomes shallow and both the exciton and trion binding energies decrease.  On the other hand, the calculated trion and exciton binding energies of a
monolayer encapsulated between two layers of varying dielectric constants increase upon decrease of the effective dielectric constant $\kappa$. This behavior is explained 
by the fact that the exciton binding energy scales as $\sim 1/(\lambda_{s}\kappa) $. The Darwin term correction lowers the relatively large trion binding energy inherited from neutral 
exciton \cite{Trushin2018}. The Berry curvature $\Omega$ has a similar effect in the trion binding energy, i.e. it decreases with increasing Berry curvature, 
similarly to the binding energy of the neutral exciton.  For $\lambda _{s}=30 $\AA, $ \Omega $=10\AA$ ^{2} $ and $ \kappa=1.55 $ 
corresponding to a SiO$ _{2} $ substrate, the trion binding energy of WS$ _{2} $ monolayer is around 35 meV, which is in excellent agreement with the value of 34 meV reported in 
Ref. \cite{Zhu2015}. Accordingly, the exciton binding energy is around 278 meV, using the same $ \lambda_{s} $, which is somewhat lower than 
experimental values of 340 meV. In order to reproduce the experimental exciton and trion binding energies of about 300 and 42meV at 7K, respectively \cite{Jadczak2017}, 
we use $ \lambda_{s}=28 $\AA\ for a  WS$ _{2} $ monolayer on SiO$ _{2} $ substrate. In similar structures but made of WSe$ _{2} $ monolayer, B$ _{1s} $= 290 meV for a 
sample placed on SiO$ _{2} $ substrate $ \varepsilon_{sub}=2.1 $ which is in agreement with reported in Ref. \cite{Raja2017} using $ \lambda_{s}=26 $\AA. A relatively 
small exciton binding energy around 160 meV has been extracted for a WSe$ _{2} $ monolayer encapsulated in hBN  \cite{Molas2018}. As one expects, the increase of 
the dielectric constant lowers the characteristic energies, while keeping 
the binding energy lower if the Berry curvature is taken into account as compared to $\Omega=0$. 
It should be noted that our calculation concern only the intra-valley trion which involves a pair of electrons from the same subband in the conduction band.

\section{Discussion and conclusions}

In conclusion, we have investigated the role of Berry-curvature corrections in the energies of symmetric intravalley charged excitons (trions). 
The analysis is based on previous work on neutral excitons \cite{Zhou2015,srivastava2015,Trushin2018}, which we have  reviewed in the first part of our paper. 

With this modified Hamiltonian, the Berry curvature changes the neutral and charged exciton binding energy, mixing and slitting angular momentum states and reorganizing the spectrum. 
The result is that the different angular momentum states for each energy level $n$ have their energies split from each other. On the other hand, 
the Darwin term leads to an energy shift depending on the  radial quantum number $n$. Furthermore, the charged trion energy is also found to exhibit a strong shift 
since it depends on exciton states, as anticipated in Eq (30). Tables \ref{tab4} and \ref{tab5} contain our complete findings  for the exciton and trion binding energies 
dependence on the Berry curvature for a range of screening lengths and relative dielectric constants of the bottom substrate. First, the decrease of the exciton and 
trion binding energies is due to the strong dielectric screening by the surrounding environment even for $\Omega\neq 0$. Second, in addition to the effect of the 
dielectric environment surrounding the sample, 
the behavior of the binding energies is also related to changes in the Berry curvature. For the same substrate (i.e. for fixed $\varepsilon_{sub}$) and by varying the screening 
length, the Berry correction acts 
dramatically for a relatively large exciton binding energy corresponding to small $\lambda_s$. Consequently, the trion binding energy shifted by about 20 meV and 
sometimes exceeds the experimental findings. 
This may explain the significant TMDs monolayer exciton and trion binding energies discrepancy using for example SiO$_{2}$ substrate, 
i.e. $\varepsilon_{sub}$=2.1, and thus $\kappa=$ 1.55.  Berry-curvature effects 
are less important for relatively week exciton and trion binding energy. Our results exhibit relatively agreement with those of Ref. \cite{Jadczak2017}. 
Similar to the findings for exciton binding energy by varying $\lambda_s$ or $\varepsilon_{sub}$, the trion binding energy reacts most sensitively to changes 
in $\Omega$. The most striking feature of the trion binding energy dependence on Berry correction is $B_{T}(\Omega)-B_{T}(0)=\left[ B_{X}(\Omega)-B_{X}(0)\right] /2$

Finally, by examining the excitonic spectrum, we have shown that the degeneracy of 2D excitonic states is lifted due to the inclusion of the Berry curvature. 
The result is a dramatic reorganization of the excitonic spectrum, 
producing thus a new distribution of trion states.


\begin{thebibliography}{99}
\bibitem{foot2}
This can be achived if we substitute $\partial \hat{H}_0/\partial \Pib \simeq \dot{\bR}$ and $\partial \hat{V}/\partial \bR \simeq -\dot{\Pib}$, in the corresponding
equations, the corrective terms being of higher order in $B$ and $\Omega_\alpha$.
\bibitem{foot4}
Notice, however, that it is strictly speaking a redundant variable that can be expressed in terms of $\rho_1$ and $\rho_2$.
\bibitem{foot5}
Corrections beyond this approximation are parabolic in the wave vector and would thus lead to terms of order $\vartheta(p^{3)}$ in HB that are also neglected in the kinetic term of the band Hamiltonian.
\bibitem{foot6}
Notice that the validity of this result does not depend on the assumption $\Omega_{e}(p)=-\Omega_{h}(p)$, which is only valid in a true two-band model. 
In this case, one needs to simply consider the exciton Berry curvature as the difference between the two, $\Omega(0)=\Omega_{e}(0)-\Omega_{h}(0)$.
\end{thebibliography}
\end{document}